\documentclass[aps,preprintnumbers, twocolumn, showpacs]{revtex4}
\usepackage{graphicx}
\usepackage{amsmath}
\usepackage{bm}
\usepackage{color}

\def\be{\begin{equation}}
\def\ee{\end{equation}}
\def\bea{\begin{eqnarray}}
\def\eea{\end{eqnarray}}

\begin{document}

\title{Wormhole geometries in Eddington-inspired Born-Infeld gravity}

\author{Tiberiu Harko$^1$}
\email{t.harko@ucl.ac.uk}
\author{Francisco S.~N.~Lobo$^{2}$}
\email{flobo@cii.fc.ul.pt}
\author{M. K. Mak$^{3}$}
\email{mkmak@vtc.edu.hk}
\author{Sergey V. Sushkov$^{4}$}
\email{sergey_sushkov@mail.ru}
\affiliation{$^1$Department of Mathematics, University College London, Gower Street,
London WC1E 6BT, United Kingdom}
\affiliation{$^2$Centro de Astronomia e Astrof\'{\i}sica da Universidade de Lisboa, Campo
Grande, Ed. C8 1749-016 Lisboa, Portugal}
\affiliation{$^{3}$Department of Computing and Information Management, Hong Kong
Institute of Vocational Education, Chai Wan, Hong Kong}
\affiliation{$^4$Institute of Physics, Kazan Federal University, Kremlevskaya Street 18,
Kazan 420008, Russia}
\date{\LaTeX-ed \today}

\begin{abstract}

Eddington-inspired Born-Infeld gravity (EiBI) gravity is a recently proposed modified theory of gravity, based on the classic work of Eddington and on Born-Infeld nonlinear electrodynamics. In this paper, we consider the possibility that wormhole geometries are sustained in EiBI gravity. We present the gravitational field equations for an anisotropic stress-energy tensor and consider the generic conditions, for the auxiliary metric, at the wormhole throat. In addition to this, we obtain an exact solution for an asymptotically flat wormhole.

\end{abstract}

\maketitle



{\it Introduction:}
An important and challenging current problem in cosmology is the perplexing fact that the Universe is undergoing an accelerating expansion \cite{cosmicacc}. The latter represents a new imbalance in the governing gravitational field equations and a possible candidate for the late-time cosmic acceleration is the introduction of a repulsive fluid dubbed ``dark energy''. Recently, by combining the Planck data in \cite{Riess}, the authors provide an equation of state parameter of dark energy given by $w=-1.24^{+0.18}_{-0.19}$, which is off  by more than the 2$\sigma $ compared to $w=-1$ \cite{Planck2}. Thus, the Planck data favours an equation of state that lies in the phantom regime, i.e., with $w<-1$. The latter equation of state consequently violates the null energy condition, and as this is the fundamental ingredient to sustain traversable wormholes \cite{Morris}, this cosmic fluid presents us with a natural scenario for the existence of these exotic geometries \cite{phantomWH}.

However, the possibility that the late-time cosmic acceleration is due to infra-red modifications \cite{Carroll:2003wy} of General Relativity (GR) has also been extensively explored in the literature.
The latter modified theories of gravity involve more general combinations of curvature invariants than the pure
Einstein-Hilbert term \cite{fRgravity}, and an interesting application in wormhole physics, is that these exotic geometries can be theoretically constructed without the presence of exotic matter \cite{modgravWH}. More specifically, the matter threading the wormhole throat may be imposed to satisfy all of the energy conditions, and it is the higher order curvature terms, which may be interpreted as a gravitational fluid, that support these nonstandard wormhole geometries \cite{Harko:2013yb}.

In the context of modified gravity, an interesting theory has been recently proposed, dubbed Eddington-inspired Born-Infeld (EiBI) gravity \cite{Deser,1}. The latter was inspired in the Eddington gravitational action \cite{4} and Born-Infeld nonlinear electrodynamics \cite{5}, and has aroused much interest. In EiBI gravity, the theory is based on a Palatini-type formulation, where the metric tensor and the connection are varied
independently. In fact, the Eddington action is coupled to matter without insisting on a purely affine action, and reduces to standard GR in vacuum, but presents a different behavior of the gravitational field in the presence of matter.

The structure of compact stars in EiBI theory has been intensively
investigated \cite{stars}, and Solar observations, big bang nucleosynthesis, and the existence of neutron stars \cite{1,stars2} have placed constraints on the value and sign of the Eddington coupling parameter $\kappa $. In particular, for cases with positive $\kappa $, an effective gravitational repulsion prevails, leading to the existence of pressureless stars, which are made of non-interacting particles which provide interesting models for self-gravitating dark matter, and increases in the mass limits of compact stars  \cite{stars2}. More recently, the density profile of pressureless dark matter in EiBI gravity was analysed \cite{Harko:2013xma}. It was shown that all the relevant astrophysical quantities can be predicted from the theory, and be directly compared with the corresponding observational parameters.

In this paper, we consider the possibility that wormhole geometries are sustained in EiBI gravity \cite{Deser,1}. More specifically, we present the gravitational field equations, consider the generic conditions at the wormhole throat and obtain an asymptotically flat wormhole solution. \\


{\it Eddington-Inspired Born-Infeld gravity:}
The action in EiBI theory is given by
\begin{eqnarray}
S&=&\frac{1}{16\pi }\frac{2}{\kappa }\int d^{4}x\left( \sqrt{-\left| g_{\mu
\nu }+\kappa R_{\mu \nu }\right| }-\lambda \sqrt{-g}\right)
   \nonumber \\
&&+S_{M}\left[
g,\Psi _{M}\right] ,  \label{1c}
\end{eqnarray}
where $\kappa $ is a parameter with the inverse dimension of the cosmological constant $\Lambda $, and $\lambda \neq 0$ is a dimensionless parameter. $R_{\mu\nu }$ is the symmetric part of the Ricci tensor and is constructed solely from the connection $\Gamma _{\beta \gamma }^{\alpha }$. The matter action $S_{M}$ depends only on the metric $g_{\mu \nu }$ and the matter field $\Psi_{M}$. For notational simplicity, the determinant of the tensor $g_{\mu \nu }+8\pi \kappa R_{\mu \nu }$ is denoted by $\left| g_{\mu \nu }+8\pi \kappa R_{\mu \nu
}\right| $ throughout the paper.

Taking into account the limit $\kappa \rightarrow 0$, the action (\ref{1c}) recovers the Einstein-Hilbert action with $\lambda =\Lambda \kappa +1$. However, in the present paper, we consider asymptotic flat solutions and hence take $\lambda =1$, so that the cosmological constant vanishes. Thus, the remaining parameter $\kappa $ plays an important role in describing the physical behavior of various cosmological and stellar scenarios.

Note that the metric $g_{\mu \nu }$ and the connection $\Gamma_{\beta \gamma }^{\alpha }$ are treated as independent fields in EiBI theory. Now, by varying the action (\ref{1c}) with respect to the connection $\Gamma_{\beta\gamma }^{\alpha }$ and with respect to the real metric $g_{\mu\nu }$ yield the following field equations
\begin{eqnarray}
q_{\mu \nu }&=&g_{\mu \nu }+\kappa R_{\mu \nu },  \label{2c}\\
q^{\mu \nu }&=& \tau \left( g^{\mu \nu }-8\pi \kappa T^{\mu \nu }\right) ,
\label{3c}
\end{eqnarray}
where the auxiliary metric $q_{\mu \nu }$ is related to the
connection by
$\Gamma_{\beta \gamma }^{\alpha }=\frac{1}{2}q^{\alpha \sigma }\left(
\partial_{\gamma }q_{\sigma \beta }+\partial _{\beta }q_{\sigma \gamma}-\partial _{\sigma }q_{\beta \gamma }\right)$, and we have defined $\tau $ and the stress-energy tensor as
$\tau =\sqrt{g/q}$, and $T^{\mu \nu }=\frac{1}{\sqrt{-g}}\frac{\delta S_{M}}{\delta g_{\mu \nu }}$, respectively.
If the stress-energy tensor $T^{\mu \nu }$ vanishes in Eq. (\ref{3c}), then the real metric $g_{\mu \nu }$ is equal to the apparent metric $q_{\mu \nu }$, and EiBi gravity is completely equivalent to GR in vacuum.

From Eqs. (\ref{2c}) and (\ref{3c}), we obtain the following relations:
$q^{\mu \sigma }g_{\sigma \nu }=\delta _{\nu }^{\mu }-\kappa R_{\nu
}^{\mu }$ and $q^{\mu \sigma }g_{\sigma \nu }=\tau \left( \delta _{\nu }^{\mu }-8\pi \kappa T_{\nu }^{\mu }\right)$, where $q^{\mu \nu }$ and $g^{\mu \nu }$ are the matrix inverse of $q_{\mu\nu }$ and $g_{\mu \nu }$ respectively, and we have used the definitions $R_{\nu }^{\mu }=q^{\mu \sigma }R_{\sigma \nu }$ and $T_{\nu }^{\mu }=T^{\mu \sigma }g_{\sigma \nu }$. By combining the above relations yields the following results:
$R_{\nu }^{\mu }=8\pi \tau T_{\nu}^{\mu}+\frac{1-\tau}{\kappa}\delta_{\nu}^{\mu}$, $R=8\pi \tau T+\frac{4(1-\tau) }{\kappa}$, where we have used $R=R_{\mu }^{\mu }$ and $T=T_{\mu }^{\mu }$.

Noting that the Einstein tensor is defined for the apparent metric $q_{\mu\nu}$, the gravitational field equation then follows immediately
\begin{equation}
G_{\nu}^{\mu }=R_{\nu}^{\mu }-\frac{1}{2} \delta_{\nu}^{\mu }R=8\pi S_{\nu}^{\mu },  \label{15c}
\end{equation}
where the apparent stress-energy tensor $S_{\nu }^{\mu }$ is defined as
\begin{equation}
S_{\nu }^{\mu }=\tau T_{\nu }^{\mu }- \left( \frac{1-\tau}{8\pi \kappa} + \frac{\tau}{2} T  \right)
\delta _{\nu }^{\mu }.  \label{16c}
\end{equation}
The stress-energy tensor $T^{\mu \nu }$ satisfies the conservation equations $\nabla _{\mu }T^{\mu \nu }=0$, where as in GR, the covariant derivative $\nabla_{\mu }$ refers to the metric $g_{\mu\nu}$. Consistently, the Bianchi identities $\nabla_{\mu }G^{\mu \nu }=0$ imply $\nabla_{\mu }S^{\mu \nu }=0$.

We will use the field equation (\ref{15c}) in the analysis outlined below. Thus, the study of the structure of wormhole geometries in the framework of EiBI theory reduces to the study of this potentially exotic source $S_{\nu }^{\mu }$.



{\it Gravitational field equations:}
Consider that the static and spherically symmetric line elements for the real metric $g_{\mu \nu }$ and the auxiliary metric $q_{\mu \nu }$ take the following forms
\begin{eqnarray}
g_{\mu \nu }dx^{\mu }dx^{\nu }&=&-e^{\nu \left( r\right) }dt^{2}+e^{\lambda
\left( r\right) }dr^{2}+f\left( r\right) d\Omega ^{2},  \label{physmetric}
     \\
q_{\mu \nu }dx^{\mu }dx^{\nu }&=&-e^{\beta \left( r\right) }dt^{2}+e^{\alpha \left( r\right) }dr^{2}+r^{2}d\Omega ^{2},  \label{9c}
\end{eqnarray}
respectively, where $\nu \left( r\right) $, $\lambda \left( r\right) $, $\beta \left(r\right) $, $\alpha \left( r\right) $ and $f\left( r\right) $ are arbitrary metric functions of the radial coordinate $r$, and we have defined $d\Omega^{2}$ as $d\Omega^{2}=d\theta^{2}+\sin ^{2}\theta d\phi ^{2}$.

The physical stress-energy tensor for an anisotropic distribution of matter is provided by
\begin{equation}
T_{\mu\nu}=(\rho+p_t)u_\mu \, u_\nu+p_t\,
g_{\mu\nu}+(p_r-p_t)\chi_\mu \chi_\nu \,,
\end{equation}
where $u^\mu$ is the four-velocity in the physical $g$-metric, $\chi^\mu$ is the unit spacelike vector in the radial direction, i.e., $\chi^\mu=e^{-\lambda/2}\,\delta^\mu{}_r$. The stress-energy components are defined in the following manner: $\rho(r)$ is the energy density, $p_r(r)$ is the radial pressure measured in the direction of $\chi^\mu$, and $p_t(r)$ is the tangential pressure measured in the orthogonal direction to $\chi^\mu$. The four-velocity of the fluid is normalized according to $u^{\mu }u^{\nu}g_{\mu\nu}=-1$. In view of the auxiliary (apparent) line element (\ref{9c}), the apparent four velocity of the fluid $v^{\mu }$ satisfies the condition $v^{\mu }v^{\nu }q_{\mu \nu }=-1$.

Note that the factor $\tau $ can be obtained from the $T_{\nu }^{\mu }$ by $\tau =\left| \delta _{\nu }^{\mu }-8\pi \kappa T_{\nu }^{\mu }\right| ^{-\frac{1}{2}}$, so that $\tau $ can be expressed in terms of physical quantities by
\begin{equation}
\tau =\left[ \left( 1+8\pi \kappa \rho \right) \left( 1-8\pi \kappa p_r\right)
\left( 1-8\pi \kappa p_t\right)^{2}\right] ^{-\frac{1}{2}}.  \label{19c}
\end{equation}

The system of the gravitational field equations describing the structure of traversable wormholes in EiBI gravity are  provided by Eq. (\ref{15c}), which yields the following non-zero components
\begin{eqnarray}
\frac{1}{ r^2} - \frac{e^{-\alpha}}{r^2} +\frac{\alpha' e^{-\alpha}}{r} = \frac{1}{2 \kappa }\left( \frac{a}{bc^2} - \frac{b}{ac^2}-\frac{2}{ab} + 2  \right)  ,
     \label{fieldeqs1}  \\
- \frac{1}{ r^2}  + \frac{e^{-\alpha}}{r^2} +\frac{\beta' e^{-\alpha}}{r} = \frac{1}{2 \kappa }\left( \frac{a}{bc^2} - \frac{b}{ac^2}+\frac{2}{ab} - 2  \right) ,  \label{fieldeqs2}
\end{eqnarray}
\begin{eqnarray}
\frac{e^{-\alpha}}{r} \left[2\beta'' r - \left(\alpha'-\beta' \right)\left(  2+\beta' r  \right)   \right]
= \frac{2}{\kappa }\left( \frac{a}{bc^2} + \frac{b}{ac^2}-2  \right) , \label{fieldeqs3}
\end{eqnarray}

The field equation (\ref{3c}) provide the following relations
\begin{eqnarray}
e^{\beta } = \frac{b c^{2}}{a}e^{\nu} , \qquad e^{\alpha }= \frac{ac^2}{b}e^{\lambda },
\qquad f=\frac{r^{2}}{ab},  \label{fieldeqs4}
\end{eqnarray}%
where we have defined the arbitrary functions $a(r) $, $b(r) $ and $c(r)$ as $a=\sqrt{1+8\pi \kappa \rho }$, $b=\sqrt{1-8\pi \kappa p_r}$, and $c=\sqrt{1-8\pi \kappa p_t}$, respectively.

In the $g$--metric the conservation of the stress-energy tensor yields the conservation equation
\begin{equation}
\frac{d\nu }{dr}=\frac{4}{r} \frac{p_t -p_r}{\rho + p_r}  -\frac{2}{p_r+\rho }\frac{dp_r}{dr}= \frac{4}{r} \frac{b^2-c^2}{a^2-b^2}+ \frac{4b}{a^{2}-b^{2}}\frac{db}{dr}.  \label{conservationeq}
\end{equation}

Note that the system (\ref{fieldeqs1})--(\ref{fieldeqs3}) is incomplete, since it consists of three equations for five unknown functions $\alpha (r)$, $\beta (r)$, $a(r)$, $b(r)$ and $c(r)$. Usually, as additional equations, one may consider two equations of state. For example, the barotropic equations of state $p_r=p_r(\rho)$ and  $p_t=p_t(\rho)$ imposes a similar equation in the $q$-metric, $b=b(a)$ and $c=c(a)$. On the other hand, instead of fixing the equation of state, one can fix two of the unknown functions using some additional reasons.


{\it Generic conditions at the wormhole throat:}
The fundamental ingredient in wormhole physics is the flaring-out condition of the throat \cite{Morris}, which through the Einstein field equation in standard GR, entails the violation of the null energy condition (NEC). It will be useful to revert to the more familiar notation currently used in wormhole physics, where we define the physical metric function as $e^{-\lambda(r)}=1-m(r)/r$, where $m(r)$ is the usual wormhole shape function \cite{Morris}. The throat is defined as $m(r_0)=r_0$, and the flaring-out condition is given by $(m'r-m)/m^2<0$.

Now, an interesting feature in EiBI gravity is the following: as the flaring-out condition imposes a restriction on the physical metric function, i.e., $\lambda' e^{-\lambda}= (m' r-m)/r^2 < 0$, through the second condition of Eq. (\ref{fieldeqs4}), i.e., $e^{\alpha }=e^{\lambda } \, ac^2/b$, one arrives at the following condition imposed on the auxiliary metric function
\begin{equation}
\lambda' e^{-\lambda}= \frac{ac^2}{b} e^{-\alpha} \left[ \alpha' - \left( \frac{ac^2}{b}\right)' \Big/\left( \frac{ac^2}{b}\right)  \right] < 0\,,
\end{equation}
which implies that
\begin{equation}
 \alpha' < \left[ \ln \left( \frac{ac^2}{b}\right)  \right]'.
\end{equation}
This latter restriction is translated as the generic flaring-out condition imposed on the auxiliary metric, in the context of EiBI gravity.

It is interesting to verify under which conditions one may impose that the normal matter threading the wormhole generically satisfies the NEC. To this effect, adding Eqs. (\ref{fieldeqs1}) and (\ref{fieldeqs2}), provides the following relationship
\begin{equation}
\frac{1}{\kappa} \left( \frac{a}{bc^2} - \frac{b}{ac^2} \right)=\left( \alpha' + \beta' \right) \frac{e^{-\alpha}}{r} .
  \label{qNEC}
\end{equation}
From Eqs. (\ref{fieldeqs4}), one obtains
$\alpha' + \beta' = \nu' + \lambda' + (c^2)'/c^2$
and substituting this into Eq. (\ref{qNEC}), one obtains
\begin{equation}
\frac{1}{\kappa} \left( \frac{a^2}{b^2} - 1 \right)=\left[\nu' + \lambda' + \frac{(c^2)'}{c^2} \right] \frac{e^{-\lambda}}{r} .
  \label{qNEC2}
\end{equation}

Taking into account Eq. (\ref{qNEC2}) and using the flaring-out condition, one arrives at the following condition
\begin{equation}
8\pi \kappa (\rho +p_r) < \frac{\kappa b^2}{r} \, \frac{(c^2)'}{c^2}\, \left( 1- \frac{m}{r} \right)\,,
\end{equation}
which is valid at the throat, or its vicinity. Now, evaluated at the throat $r=m(r)=r_0$, and considering that the factor $(c^2)'/c^2$ is finite, then on readily verifies that the NEC is violated, i.e., $\rho+p_r<0$, so that as in standard GR one also needs exotic matter to support these exotic geometries in EiBI gravity. However, a subtle feature is that considering that the factor scales as $(c^2)'/c^2 \propto (1-m/r)^{-1}$, so that it diverges as $r\rightarrow r_0$, then the possibility arises that $0<\rho+p_r < K$. Thus, for this specific case, one verifies the possibility that the NEC is satisfied by the matter threading the wormhole.


{\it Specific exact solution:}
Consider the specific choice of $a(r)b(r)=1$, and using the third relation of Eqs. (\ref{fieldeqs4}), one obtains $f(r)=r^2$. Now, the definitions of $a(r)$ and $b(r)$ yield the following equation of state
\be
p_r(r)=\frac{\rho(r)}{1+8\pi \kappa \rho(r)}.
 \label{radialp}
\ee

Conveniently reorganizing the gravitational field equations (\ref{fieldeqs1})-(\ref{fieldeqs4}), we deduce the system of equations
\begin{eqnarray}
\beta' - \alpha' + \frac{2}{r^2} - \frac{2e^{\alpha}}{r^2}=0  ,  \label{fieldeqs1b} \\
%
\frac{1}{\kappa }\left( \frac{a^2}{c^2} - \frac{1}{a^2c^2}\right) =(\beta' + \alpha')\frac{e^{-\alpha}}{r},  \label{fieldeqs2b}
\end{eqnarray}
\begin{eqnarray}
\frac{2}{\kappa }\left( \frac{a^2}{c^2} + \frac{1}{a^2c^2}-2  \right) = \frac{e^{-\alpha}}{ r}
\left[2\beta'' r - \left( \alpha'-\beta' \right)\left(  2+\beta' r  \right)   \right] , \label{fieldeqs3b}
\end{eqnarray}
\begin{eqnarray}
e^{\beta } = e^{\nu} \frac{c^{2}}{a^2}, \qquad e^{\alpha }=e^{\lambda } a^2c^2,
  \label{fieldeqs4b}
\end{eqnarray}%
respectively.

To close the system of equations, and obtain specific exact solutions, we may adopt several strategies. For simplicity,
consider $\beta=0$, so that Eq. (\ref{fieldeqs1b}) yields
\begin{equation}
e^{\alpha(r)}=\left(1-\frac{C_1}{r^2}  \right)^{-1},
\end{equation}
where $C_1$ is a constant of integration.
Equations (\ref{fieldeqs2b})-(\ref{fieldeqs3b}) provide the equation of state $c^2(r)=a^2(r)$, i.e., $p_t(r)=-\rho(r)$, and one readily finds the solution
\begin{equation}
a^4=\frac{1}{1+2\kappa C_1/r^4}\,.
  \label{defa}
\end{equation}

Taking into account Eqs. (\ref{fieldeqs4b}), we find the following solutions for the physical metric functions
\begin{equation}
\nu (r)=0, \qquad e^{\lambda(r)}=\frac{1+2\kappa C_1/r^4}{1-C_1/r^2}.
\label{shape1}
\end{equation}
where the constant is defined as $C_1=r_0^2$ in order to have a wormhole throat at $r=r_0$.

As mentioned above, the flaring-out condition at the throat places the restriction on the shape function, so that from the metric function given by Eq. (\ref{shape1}), we have
\begin{equation}
\lambda' e^{-\lambda}= - \frac{2r_0^2r(r^2-2\kappa r_0^2+4\kappa r^2)}{(r^4+2\kappa r_0^2)^2}\,,
\end{equation}
which yields $\lambda' e^{-\lambda}|_{r_0}=-2r_0/(r_0^2+2\kappa)<0$ at the throat, and thus satisfies the flaring-out condition at the throat.

Thus, one finally arrives at the asymptotically flat physical metric given by
\begin{eqnarray}
ds^2=-dt^2+ \left( \frac{1+2\kappa r_0^2/r^4}{1-r_0^2/r^2} \right) \,dr^2+r^2\, d\Omega^2 .
\end{eqnarray}
It is interesting to note that in the limiting case $\kappa \rightarrow 0$, this line element tends to the phantom massles scalar field solution found by Ellis and Bronnikov \cite{EllisBronn}.
The stress-energy tensor profile is given by the energy density
\begin{eqnarray}
\rho(r)&=&\frac{1}{8\pi \kappa}\left( \frac{1}{\sqrt{1+2\kappa r_0^2/r^4}} -1\right),
\end{eqnarray}
which can be readily deduced from Eq. (\ref{defa}); the radial pressure is given by Eq. (\ref{radialp}), while the tangential pressure is provided by $p_t=-\rho$.
Note that for this case, the energy density and the radial pressure are negative throughout the spacetime, and so violate the NEC, while the tangential pressure is positive $\forall r$. The energy density attains a minimum value at the throat, $\rho(r_0)=-(1-1/\sqrt{1+2\kappa /r_0^2})/(8\pi \kappa)$ and falls to zero at spatial infinity, i.e., $\rho \rightarrow 0$ at $r \rightarrow \infty$ ($p_r$ and $p_t$ have a similar behaviour at spatial infinity).


{\it Conclusion:}
In this paper, we have considered the possibility that wormhole geometries are sustained in EiBI gravity. We considered the generic conditions at the wormhole throat and obtained an exact solution for an asymptotically flat wormhole, that reduces to the phantom massles scalar field geometry found by Ellis and Bronnikov \cite{EllisBronn}. In this context, wormholes have been modelled by so-called ghost or phantom scalar fields, and the stability issue has been extensively explored in the literature. In particular, it has been shown that massless non-minimally coupled scalar fields are unstable under linear \cite{linearstability} and nonlinear perturbations \cite{nonlinearstability}.

In fact, two serious difficulties seem to stand out, namely, $(i)$ instabilities arise at boundary surfaces separating ghost and normal fields, which generally transform these surfaces into singular ones \cite{stable1}; $(ii)$ and ghost fields are extremely problematic at the quantum level. In the latter feature, the negative kinetic term leads to the possibility that the energy density may become arbitrarily negative for high frequency oscillations \cite{stable2}. In this context, it would be interesting to apply a stability analysis to the EiBI wormholes geometries considered in this paper, and work along these lines is presently underway.


\textit{Acknowledgments.} FSNL acknowledges financial support of the Funda%
\c{c}\~{a}o para a Ci\^{e}ncia e Tecnologia through the grants
CERN/FP/123615/2011 and CERN/FP/123618/2011. SVS acknowledges financial
support of the Russian Foundation for Basic Research through grants No.
11-02- 01162 and 13-02-12093 .

\label{sect3}


\begin{thebibliography}{99}

\bibitem{cosmicacc}
A. Grant {\it et al},
Astrophys. J. {\bf 560} 49-71 (2001);
%
S. Perlmutter, M. S. Turner and M. White,
Phys. Rev. Lett. {\bf 83} 670-673 (1999).

\bibitem{Riess}   A. G. Riess, L. Macri, S. Casertano,  et al.,  Astrophys. J. {\bf 730}, 119 (2011).

\bibitem{Planck2} P. A. R. Ade et al., Planck 2013 results. XVI, arXiv: 1303.5076
[astro-ph] (2013).

\bibitem{Morris}
M. Morris and K.S. Thorne,
Am. J. Phys. \textbf{56}, 395 (1988).


\bibitem{phantomWH}
S.~V.~Sushkov,
  Phys.\ Rev.\ D {\bf 71}, 043520 (2005);
  F.~S.~N.~Lobo,
  Phys.\ Rev.\ D {\bf 71}, 084011 (2005).

\bibitem{Carroll:2003wy}
  S.~M.~Carroll, V.~Duvvuri, M.~Trodden and M.~S.~Turner,
  Phys.\ Rev.\ D {\bf 70}, 043528 (2004)

\bibitem{fRgravity}
S.~'i.~Nojiri and S.~D.~Odintsov,
  Int.\ J.\ Geom.\ Meth.\ Mod.\ Phys.\  {\bf 4}, 115 (2007);
A.~De Felice and S.~Tsujikawa,
  Living Rev.\ Rel.\  {\bf 13}, 3 (2010);
%
 F.~S.~N.~Lobo,
  arXiv:0807.1640 [gr-qc].
  S. Capozziello, M. De Laurentis,
  Phys.\ Rept.  {\bf 509}, 167  (2011);
  S.~'i.~Nojiri and S.~D.~Odintsov,
  Phys.\ Rept.\  {\bf 505}, 59 (2011).

\bibitem{modgravWH}
F.~S.~N.~Lobo, M.~A.~Oliveira,
Phys.\ Rev.\ \textbf{D80}, 104012 (2009); 
F.~S.~N.~Lobo,
Class.\ Quant.\ Grav.\ \textbf{25}, 175006 (2008);
F.~S.~N.~Lobo,
Phys.\ Rev.\ \textbf{D75}, 064027 (2007). 
N.~M.~Garcia, F.~S.~N.~Lobo,
Phys.\ Rev.\ \textbf{D82}, 104018 (2010); 
%
N.~Montelongo Garcia, F.~S.~N.~Lobo,
Class.\ Quant.\ Grav.\ \textbf{28}, 085018 (2011).

\bibitem{Harko:2013yb}
  T.~Harko, F.~S.~N.~Lobo, M.~K.~Mak and S.~V.~Sushkov,
  Phys.\ Rev.\ D {\bf 87}, 067504 (2013).

\bibitem{Deser}
S. Deser and G. W. Gibbons, Class. Quant. Grav. \textbf{15},
L35 (1998).

\bibitem{1} M. Ba\~{n}ados, P. G. Ferreira,
Phys. Rev. Lett. \textbf{105}\textit{,} 011101
(2010).

\bibitem{4} A. S. Eddington, \textit{The mathematical theory of relativity},
(Cambridge University Press, Cambridge, 1924).

\bibitem{5} M. Born, L. Infeld, \textit{Foundations of the new field theory},%
\textit{\ }Proc. R. Soc. A \textbf{144}, 425-451 (1934).


\bibitem{stars} T. Delsate, J. Steinhoff,
Phys. Rev. Lett. \textbf{109}, 021101 (2012);
%
P. Pani, T. P. Sotiriou,
Phys. Rev. Lett. \textbf{109}, 251102 (2012);
%
Y.-H. Sham, P. T. Leung, L.-M. Lin,
Phys. Rev. D \textbf{87}, 061503(R) (2013);
%
Y.-H. Sham, L.-M. Lin, P. T. Leung,
Phys. Rev. D \textbf{86}, 064015 (2012);
%
P. Pani, T. Delsate, V. Cardoso,
Phys. Rev. D \textbf{85}, 084020 (2012);
%
  T.~Harko, F.~S.~N.~Lobo, M.~K.~Mak and S.~V.~Sushkov,
  arXiv:1305.6770 [gr-qc].

\bibitem{stars2}
P. Pani, V. Cardoso, T. Delsate,
Phys. Rev. Lett. \textbf{107}\textit{,} 031101 (2011);
%
Phys. Rev. D \textbf{85}, 104053 (2012);
%
I. L. J. Casanellas {\it et al}, Astrophys. J. \textbf{745}, 15 (2012).

\bibitem{Harko:2013xma}
  T.~Harko, F.~S.~N.~Lobo, M.~K.~Mak and S.~V.~Sushkov,
  arXiv:1305.0820 [gr-qc].
  


\bibitem{EllisBronn}
H. Ellis, J. Math. Phys. {\bf 14}, 104 (1973);
K.A. Bronnikov, Acta Phys. Polonica {\bf B 4}, 251 (1973).

\bibitem{linearstability}
K.A. Bronnikov and S. Grinyok, Grav. Cosmol. 7, 297 (2001); F. S. Guzman and O. Sarbach, Class. Quant. Grav. 26, 015010 (2009).

\bibitem{nonlinearstability}
J. A. Gonzalez, F. S. Guzman and O. Sarbach, Class. Quant. Grav. 26, 015011 (2009).

\bibitem{stable1}
K. A. Bronnikov and S. V. Grinyok, Grav. Cosmol. 10, 237 (2004)

\bibitem{stable2}
S. V. Sushkov and Y. -Z. Zhang, Phys. Rev. D 77, 024042 (2008).

\end{thebibliography}
\end{document}